\documentclass[aps,prx,superscriptaddress,showpacs,twocolumn]{revtex4-1} 
\usepackage{units}
\usepackage{amsmath}
\usepackage{amssymb}
\usepackage{graphicx}
\usepackage{bm}
\usepackage{multirow,microtype,color,relsize,ulem}

\newcommand{\be}{\begin{equation}}
\newcommand{\ee}{\end{equation}}

\newcommand{\PT}{$\mathcal PT$}

\begin{document}
\title{Robustness versus sensitivity in non-Hermitian topological lattices probed by pseudospectra}
	\author{I. Komis}
	\affiliation{ITCP-Physics Department, University of Crete, Heraklion, 71003, Greece}
	\affiliation{Institute of Electronic Structure and Laser (IESL) – FORTH, Heraklion, 71110, Greece}
	\author{D. Kaltsas}
	\affiliation{ITCP-Physics Department, University of Crete, Heraklion, 71003, Greece}
		\author{S. Xia}
	\affiliation{The MOE Key Laboratory of Weak-Light Nonlinear Photonics, TEDA Applied Physics Institute and School of Physics, Nankai University, Tianjin 300457, China}
	\author{H. Buljan}
	\affiliation{Department of Physics, Faculty of Science, University of Zagreb, Bijenička c. 32, 10000 Zagreb, Croatia}
	\affiliation{The MOE Key Laboratory of Weak-Light Nonlinear Photonics, TEDA Applied Physics Institute and School of Physics, Nankai University, Tianjin 300457, China}
	\author{Z. Chen}
	\affiliation{The MOE Key Laboratory of Weak-Light Nonlinear Photonics, TEDA Applied Physics Institute and School of Physics, Nankai University, Tianjin 300457, China}
	\affiliation{Department of Physics and Astronomy, San Francisco State University, California 94132, USA}
	\author{K. G. Makris}
	\affiliation{ITCP-Physics Department, University of Crete, Heraklion, 71003, Greece}
	\affiliation{Institute of Electronic Structure and Laser (IESL) – FORTH, Heraklion, 71110, Greece}

\date{\today}

\begin{abstract}
Non-Hermitian topological systems simultaneously posses two antagonistic features: ultra sensitivity due to exceptional points and robustness of topological zero energy modes, and it is unclear which one prevails under different perturbations. We study that question by applying the pseudospectra theory on the prototypical non-Hermitian SSH (NHSSH) lattice. Topological modes are robust with respect to chiral perturbations and sensitive to parity-time (\PT) symmetry preserving perturbations. In fact, the chiral symmetry exactly at the exceptional point leads to the suppression of its sensitivity, corresponding to a lower order exceptional point. However, counterintuitively, they are most sensitive with respect to unstructured perturbations, leaving the fingerprint  of the pertinent higher order exceptional (HEP) point.

\end{abstract}

\maketitle

\textit{Introduction.}--Non-Hermitian Hamiltonians in classical and quantum physics \cite{EP1} describe the dynamics of open systems under the influence of dissipation and/or amplification. One of their intriguing characteristics is the existence of unique non-Hermitian degeneracies \cite{EP2} the so-called exceptional points (EPs) \cite{EP3}, where two or more eigenvalues and eigenvectors coalesce for a particular value of the system’s parameter \cite{EP4,EP5}, forming thus a higher order exceptional point (HEP). Motivated by the recent introduction of the concept  of parity-time (\PT)-symmetry \cite{Bender1,Bender2,Bender3} in optics \cite{PT1,PT2,PT3,PT4,PT5}, where the spatial mixing of gain and loss is physically accessible, the new area of non-Hermitian photonics \cite{PT6,PT7,PT8,PT9,PT10,PT11,PT12,PT13,PT14,PT15,PT16,PT17,PT18} has emerged \cite{Rev1,Rev2,Rev3,Rev4,Rev5}. A plethora of experimental realizations of photonic devices that operate around the HEPs is evident, and is mainly based on the enhanced response of the system around such degeneracies. Among the most impressive experiments are these related to ultra sensitive sensors \cite{PT17} and non-Hermitian gyroscopes \cite{PT18}. 

On the other hand, topological photonics \cite{Zhigang1,Marin,Rechtsman,Rechtsman2,Moti,Rechtsman3,Zhigang2} relies on the key property of topological protection of the zero eigenstate. Especially in two dimensions, such an effect leads to transport of optical waves along the edge of a photonic topological insulator, even under the presence of strong external perturbations. Most of studies are so far devoted to Floquet systems with broken time reversal symmetry that induce effective pseudomagnetic fields \cite{Rechtsman}. 

Recently however, a new frontier of non-Hermitian topological photonics \cite{Rechtsman4,Migel1,Migel2,Longhi,Haldane} has emerged, based on the synergy between the two aforementioned areas. This new direction has led to an explosion of  theoretical and experimental results that exploit the existence of chiral and non-Hermitian symmetries on the same lattice.  Among the recent experiments that define this field is the demonstration of $\mathcal{PT}$-symmetry breaking in a non-Hermitian Su-Schrieffer-Heeger (NHSSH) lattice \cite{SSH,PT-topo}, the topological insulator lasers \cite{Migel1,Migel2} and the non-Hermitian Haldane lattice \cite{Haldane}. Nonlinearity also plays an important role and provides a new degree of freedom, as a relevant recent experiment demonstrated \cite{Science}. In fact, it allows us to locally control not only the real part of the index modulation but the imaginary part as well. In such systems the inclusion of gain and loss elements make the topological lattice non-Hermitian, and thus extends the physics of topological insulators to the complex domain, with no analog whatsoever in condensed matter physics. Since most of previous concepts of topological physics have been built on the assumption of conservation laws, with underlying Hermitian operators, it means that one has to derive everything from first principles by incorporating the intricate properties of non-Hermitian algebra. Indeed, Zak phases, Chern numbers, bulk-edge correspondence and all relevant topological quantities must be redefined through the prism of non-Hermitian physics, leaving thus many open questions for further investigation \cite{Ueda1,Ueda2}. 

\begin{figure*}
    \centering
    \includegraphics[width=0.8\textwidth]{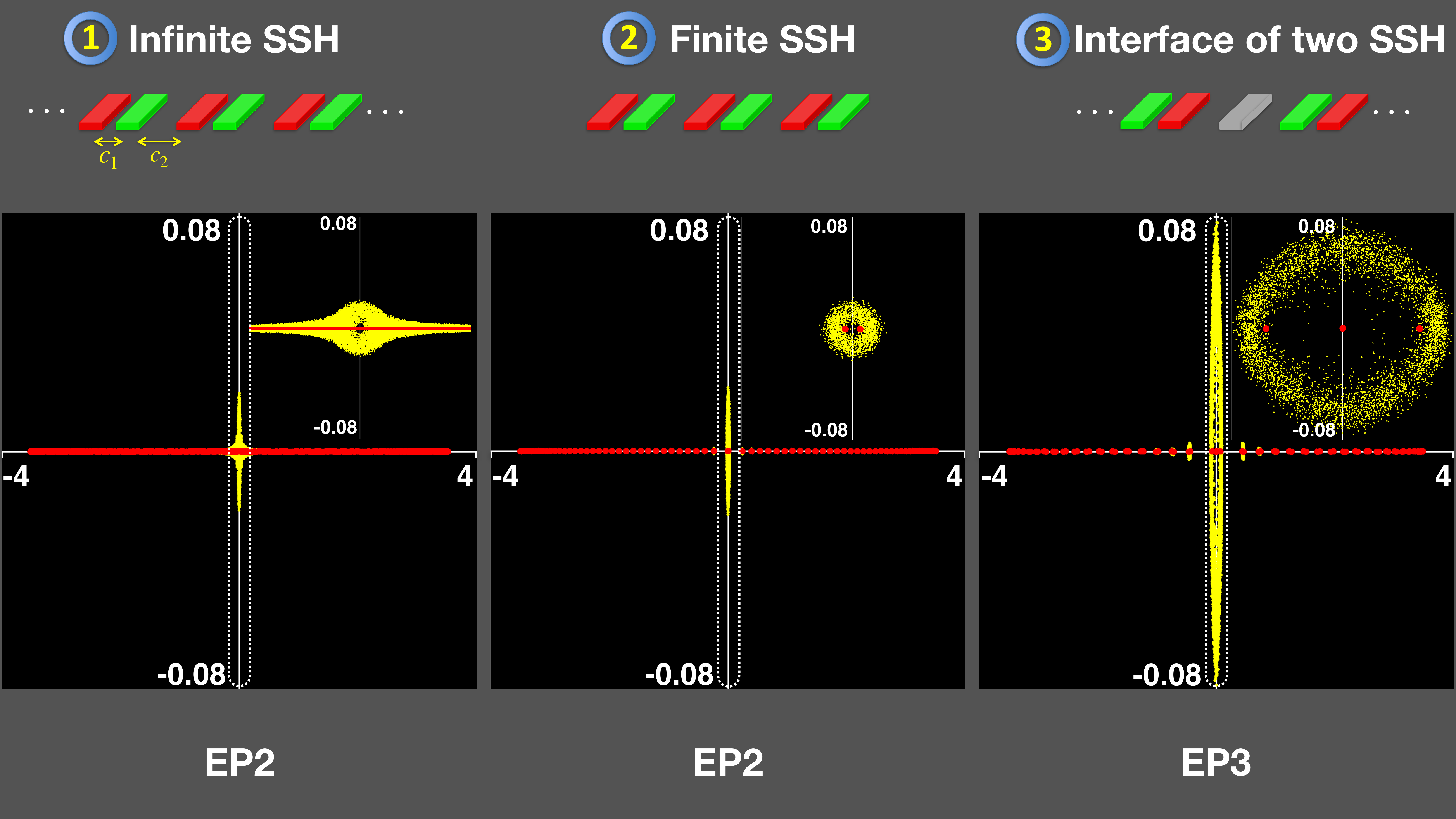}
    \caption{Complex pseudospectra of the three NHSSH lattices: infinite (left), finite (center, $N=80$) for $\gamma = 2$, and interface (right, $N=81$) for $\gamma = 2.0159$ and $\gamma_0 = 0$. In all cases the coupling constants are $c_1=3$ and $c_2=1$. On the top row the red, green and gray colors illustrate the gainy, lossy, and neutral waveguide, respectively. On the middle row, the spectra (red dots) and corresponding 0.0001-pseudospectra (100 realizations-yellow dots), for the three lattices are presented on the complex plane. Insets show a magnified view of the white-dashed area (for 1000 realizations). We have to note here that in the interface lattice, the central red dot at $[0,0]$ consists of three eigenvalues. We do not show the edge states of the finite SSH, which are located at $[0, \pm 2i]$ and have perturbations of the order of half the $\varepsilon$. On the bottom row, the order of the exceptional point that each lattice exhibits is shown.}
    \label{SSH}
\end{figure*}

In this context of non-Hermitian topological photonics, we examine the interplay between robustness (due to topology) and sensitivity (due to non-Hermiticity) in the prototypical system of an NHSSH lattice around the underlying HEPs \cite{Science}. In order to systematically examine the lattice sensitivity, we provide a mathematical framework ideal for non-Hermitian systems, that is the complex and structured pseudospectra theory \cite{TrefethenBook,Trefethen1,Trefethen2,Trefethen3}.
In particular, we consider three different lattices, namely an infinite, a finite and a hetero structure of two NHSSH lattices (see Fig.~\ref{SSH}). The intricate relation between the lattice’s symmetry and the symmetries of the global perturbations is revealed based on pseudospectra \cite{TrefethenBook,Trefethen1}. We find that for chiral structured perturbations, the topological zero state is indeed robust below the underlying EPN. Furthermore, we find that exactly at EP the chiral symmetry leads to a suppressed sensitivity that corresponds to an EP(N-1).  Counterintuitively, the zero states turn out to be most sensitive for unstructured complex perturbations, revealing the order of the pertinent exceptional point. At last we consider the lattice’s sensitivity due to single-site local perturbations of the interface channel, something that has been experimentally demonstrated using optical photorefractive nonlinearity \cite{Science}.


\textit{Non-Hermitian SSH lattices.}--Our starting point is the prototypical SSH model, in the context of coupled mode theory. More specifically, the lattices that we consider are schematically depicted in Fig.~\ref{SSH}. For the infinite lattice, the Hamiltonian in $k$-space reads,
$H^{\text{inf}}(k) = (c_2 + c_1\text{cos}k)\cdot \sigma_x + c_1\text{sin}k\cdot \sigma_y +i\gamma \cdot \sigma_z$
where, $\sigma_x,\sigma_y$, $\sigma_z$ are the Pauli matrices and $k$ is the Bloch wavenumber in the first Brillouin zone. The coupling constants are denoted by $c_1$ and $c_2$ for intra- and inter-cell coupling, respectively. The global gain-loss amplitude of each waveguide channel is described by the parameter $\gamma$, and thus making the whole system non-Hermitian. For the finite lattice the Hamiltonian matrix elements are: $H^{fin}_{nm} = \delta_{n+1,m} \cdot c_{(n \cdot {\text{mod}}2)+1} + \delta_{n,m+1} \cdot c_{(m \cdot {\text{mod}}2)+1} +\delta_{nm} \cdot i\gamma_n$,
where $\gamma_n=(-1)^n\gamma$, and $n,m = 1,...,N$. Regarding the last case of an interface SSH lattice, we consider two SSH chains that are connected with an extra channel at the interface that has a tunable gain/loss amplitude $\gamma_0$.  In such a case the Hamiltonian matrix elements are given by the last expression with the only difference that $n,m = 1,...,N/2, N/2+2,...,N+1$ with even $N$ and $H_{\frac{N}{2} + 1,\frac{N}{2} + 1} = i\gamma_0$.

In all the above three cases, the associated matrices are non-Hermitian and symmetric meaning $H^{T}=H$, unlike the Hatano-Nelson problems \cite{Hatano}. Thus their eigenvalue spectrum ($\{\lambda_n\}$) is in general complex (Supplementary information). Depending on the value of the global gain-loss amplitude $\gamma$, we find that the first two lattices exhibit an EP2 and the interface lattice an EP3.

\textit{Complex unstructured Pseudospectra.}--In the vast majority of previous studies, the sensitivity of a non-Hermitian system under external perturbations was analyzed using the semi-analytical techniques based on perturbation theory, for systems of usually only a small number of waveguides. For lattices of our type such an approach is rather problematic and is not easily applicable. Therefore we introduce an alternative and general computational framework based on \textit{pseudospectra} \cite{TrefethenBook}. The so-called geometrical spectrum or pseudospectrum is a systematic mathematical way to study the sensitivity of a matrix/operator on external perturbations, without relying on perturbation theory. It has been extensively used in the context of fluid mechanics \cite{Trefethen1,Trefethen2,Trefethen3} non-normal networks \cite{networks}, and transient growth physics \cite{growth1,growth2,growth3,growth4}. However, it is largely unknown in optics despite being ideal for studying non-Hermitian systems \cite{pseudo1,pseudo2,pseudo3,pseudo4}. For Hermitian matrices, the spectrum and the pseudospectrum are almost identical, whereas for non-Hermitian could be significantly different. The measure of how different the two spectra might be, depends on the degree of the non-orthogonality of the corresponding eigenmodes. Thus the pseudospectrum of a non-normal matrix provides us with complete information, beyond the conventional spectrum. The most basic definition of the $\varepsilon$-pseudospectrum of a non-Hermitian matrix $H$ , with $\sigma(H)$-spectrum, is the union of all spectra of the matrices $H + E$, where $E$ is a full complex random matrix (with respect to its matrix elements), with $||E|| < \varepsilon$. More specifically, $\sigma_\varepsilon (H)$ is the set of $z \in \mathbb{C}$ such that $z \in \sigma (H + E)$ for some $E \in \mathbb{C}^{N \times N}$ with $||E|| < \varepsilon$. In particular, $E = \varepsilon \frac{\mbox{E}}{\left\Vert \mbox{E} \right\Vert}$, where, $\varepsilon$ defines the perturbation strength and the matrix E is the perturbation matrix before the normalization. Our results for the three NHSSH lattices, are shown in Fig.~\ref{SSH}. The red dots represent the spectrum of the non-Hermitian system and the yellow dots the corresponding pseudospectra on the complex plane. The global gain-loss amplitude is close and below the EP2 for the infinite and finite lattices, as well as, below the EP3 for the interface lattice. As we can see the three spectra are entirely real (unbroken \PT-symmetry regime) and thus lying on the real axis (the gap is open but small). However, the corresponding pseudospectra are extended on the complex plane, and their size is related to the sensitivity of the lattice. In particular, the eigenstates close to the edges of the gap are most sensitive and these away from that gap are robust. The non-orthogonality of the eigenmodes close to the gap is significantly higher than the rest of the eigenstates. From the complex eigenvalue bifurcation curves vs $\gamma$ (see SI), we can identify the modes that coalesce and form the EP2 for the infinite and finite lattices and the EP3 for the interface lattice. The geometrical size of the main lobe of the complex pseudospectrum as a function of the perturbation strength $\varepsilon$, has a square-root and cubic-root dependence, that are characteristics of the EP2 and EP3, respectively. Another conclusion we can draw, is that as we can see, in terms of the bulk-edge correspondence, the sensitivity is similar for both finite and infinite lattices. This is not the case for asymmetric ally coupled lattices (like the Hatano-Nelson model \cite{Hatano}), because of the non-Hermitian skin effect \cite{Ueda3}.\\
\textit{Structured Pseudospectra}--Since the applied perturbations are complex and applied everywhere, even in the zero entries of the $H$-matrix, we would like to examine more realistic and experimentally relevant perturbations. Such perturbations are called structured perturbations and they define the \textit{structured pseudospectrum} \cite{TrefethenBook}, which is ideal for studying the sensitivity of our NHSSH lattice. We define the structured pseudospectrum $\sigma^{str}_\varepsilon$ of the Hamiltonian $H$, as $\sigma^{str}_\varepsilon (H)\equiv \bigcup_{j=1,E-structured,||E_j|| < \varepsilon}^{s}\sigma (H+E_j)$, where $s$ is the number of different realizations of the structured perturbations. The main difference from the unstructured pseudospectra is that the matrix $E$ is not full but has a particular structure that stems from the physics of the problem. For example if the external perturbations are applied only on the index modulation, then $E$ is diagonal. If they are applied to the coupling coefficients, then the $\pm 1$-diagonals are non-zero. In particular, we consider perturbations on the coupling coefficients i.e., the elements of the matrix are $\mbox{E}_{nm} = \delta_{n+1,m}\cdot \epsilon_n + \delta_{n,m+1}\cdot \epsilon_m ^*$, diagonal perturbations with $\mbox{E}_{nm} = \delta_{n,m}\cdot \epsilon_n$, and combination of the previous two with $\mbox{E}_{nm} = \delta_{n,m}\cdot \epsilon_m + \delta_{n+1,m}\cdot \epsilon_n + \delta_{n,m+1}\cdot \epsilon_m ^*$. For the case of the interface lattice we perturb the coupling constants as follows, $\mbox{E}^{int}_{nm} =\delta_{n+1,m} \cdot \epsilon_{(n \cdot {\text{mod}}2)+1} + \delta_{n,m+1} \cdot \epsilon_{(m \cdot {\text{mod}}2)+1}$. The complex numbers $\epsilon$ are drawn from the standard normal distribution.

\begin{figure}
    \centering
    \includegraphics[width=0.45\textwidth]{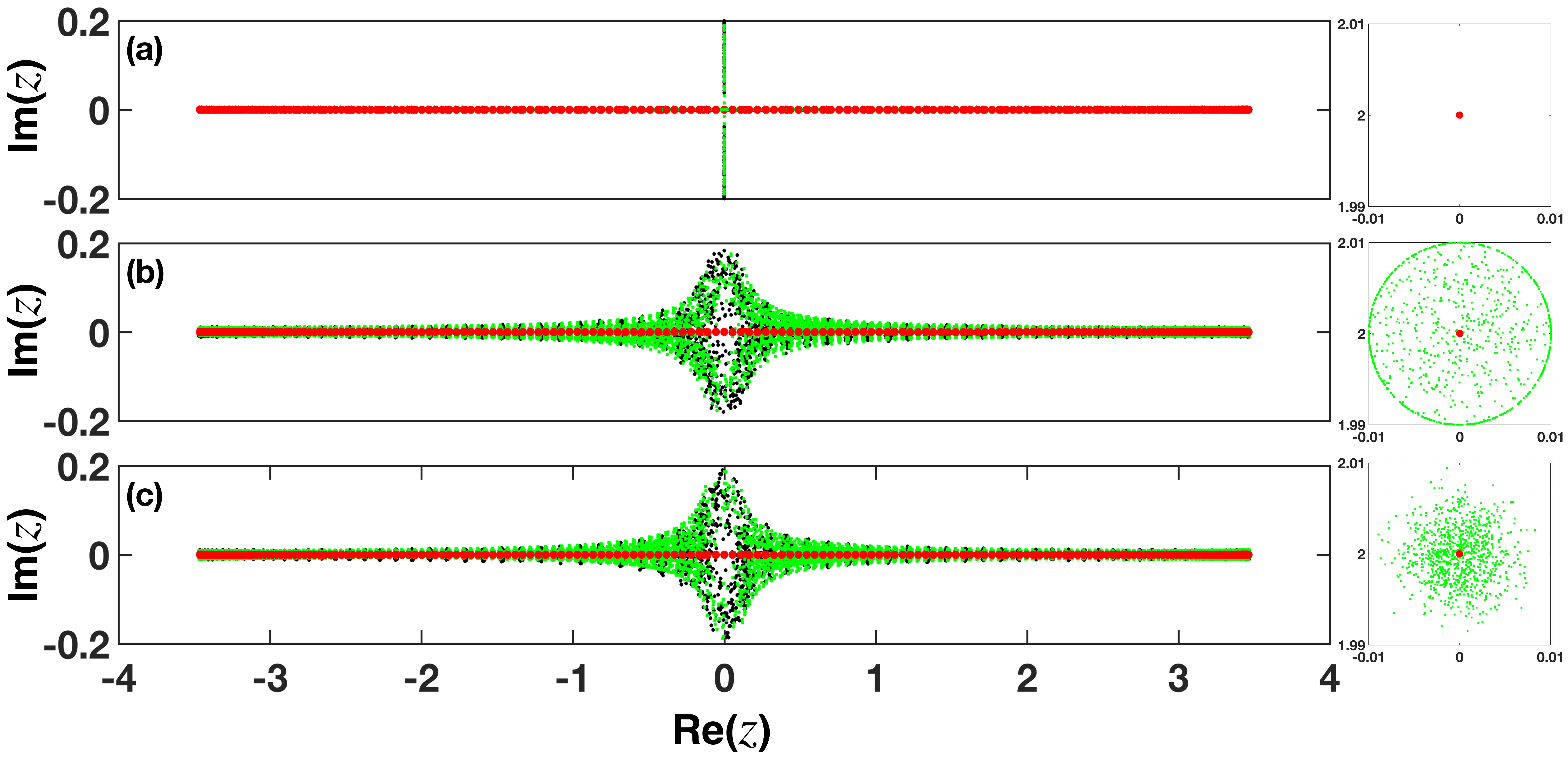}
    \caption{Structured pseudospectra $\sigma^{str}_{0.01} (H), s=60$ of infinite (black dots) and finite (green dots) NHSSH lattices. The spectrum (red dots) in all cases  is also plotted for comparison in the complex plane. (a,b,c) correspond to off-diagonal (on coupling constants), on-diagonal and combined perturbations, respectively. For each type of perturbations one of the edge states of the finite lattice is shown in right column with $s=1000$ realizations.}
   \label{inf_fin}
\end{figure}


\textit{Structured pseudospectra of the infinite and finite lattices}--Let us now consider, the infinite and finite NHSSH. For these lattices we have calculated three different types of structured pseudospectra $\sigma^{str}_{0.01} (H)$, that are shown in Fig.~\ref{inf_fin}. In particular, Figs.~\ref{inf_fin}(a,b,c) correspond to coupling, diagonal and combined perturbations, respectively. In terms of the EP's sensitivity, both approaches give us similar results, something that is expected \cite{TrefethenBook} for symmetric matrices. The difference from the complex pseudospectra (Fig.~\ref{SSH}) is that the structured perturbations on the couplings reveal the topological robustness of the edge states, as is evident from Fig.~\ref{inf_fin}(a).

\begin{figure}
    \centering
    \includegraphics[width=0.45\textwidth]{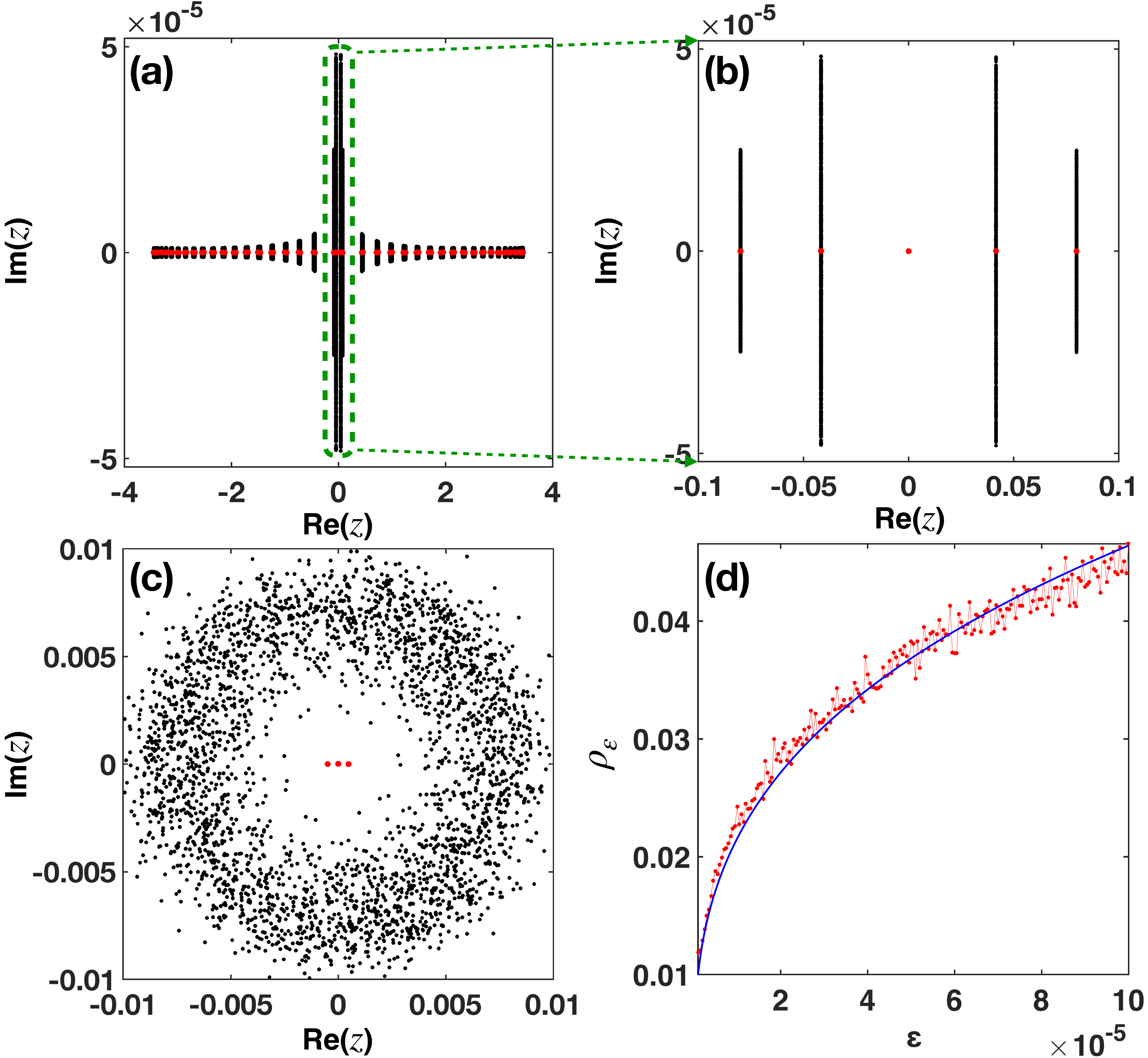}
    \caption{Structured pseudospectra $\sigma^{str}_{10^{-6}} (H)$ of the interface NHSSH lattice. (a) We include only chiral complex perturbations on the $\pm 1$ diagonals for $s=1000$ realizations, and global gain/loss amplitude $\gamma = 2.0155$ (below the EP3). As we can see the zero mode indeed remains robust. (b) Magnified view of the selected area (green dashed line) which corresponds to the five modes closer to the origin of the complex plane. The size of the gap is $\approx 0.1$. (c) Complex diagonal perturbations  for $s=1000$ realizations and  $\gamma = 2.0159293$. (d) Pseudospectral radius (red line) of (c) as a function of $\varepsilon \in [10^{-6}, \;\; 10^{-4}]$. The blue line is $\approx \varepsilon^{1/3}$ and shown for comparison.}
    \label{structured}
\end{figure}

\textit{Structured pseudospectra of the interface lattice}--Now we study the effect of different structured perturbations on the spectrum of the interface lattice. Let us start with chiral complex perturbations that physically correspond to changes on the coupling coefficients $c_1, c_2$. Since they respect the chiral symmetry of the lattice, we expect the zero-mode to be robust to such external perturbations. In order to systematically examine such a problem, we calculate the corresponding structured pseudospectrum. This time the perturbation matrix is again of size $N\times N$ but with random complex elements that respect chiral symmetry. Our results are shown in Figs.~\ref{structured}(a,b). In particular, we plot on the complex plane the eigenvalue spectrum of the lattice (red dots) and the corresponding structured pseudospectrum (black dots). As we can see, it is indeed true that the zero-mode is topologically protected, since the size of the corresponding pseudoeigenvalue cloud (black dots) is zero. The rest of the supermodes of the problem are sensitive to coupling perturbations especially for those close to the band gap edges. This is the case for small but still open gap below the EP3. Interestingly enough, for gain-loss amplitude exactly at the EP3, the gap is zero and the zero mode is not topologically protected any more. In fact, the sensitivity of lattice is not that of an EP3 (as one may expect) but that of an EP2, as a direct outcome of the chiral symmetry of the applied perturbation (see SI). 

Let us now examine the effect of diagonal perturbations on the interface lattice. Our results are presented in Figs.~\ref{structured}(c,d), for complex diagonal perturbations. The global gain-loss amplitude is close to the EP3 of the structure, and as we can see from the Fig.~\ref{structured}(c), the geometric size of the corresponding pseudospectrum is of the order of $\varepsilon ^{1/3}$, as is expected from Lidskii perturbation theory of Jordan matrices, for very small perturbation strength \cite{Davies}. Our results for perturbations both on the three main diagonals are similar to that of Fig.~\ref{structured} as shown in the SI. Under these type of perturbations it is clear that the topological robustness of the zero mode does not survive, and non-Hermitian sensitivity determines the behavior of our lattice close to the EP3. The fact that the perturbations are real or complex doesn't apparently affect their sensitivity. The associated pseudospectra have the same size. The only difference is the orientation of their extend to the complex plane (see SI). The size of the pseudospectrum can be quantitatively described by the pseudospectral radius $\rho_\varepsilon$ \cite{TrefethenBook}, which is defined here locally as $\rho_\varepsilon\equiv \mbox{max}_{z \in B}|z|$, with the $z$ belonging on the subset $B \subset \sigma_\epsilon(H)$ of interest (here the pseudoeigenvalue cloud that corresponds to the gap). In Fig.~\ref{structured}(d) we calculate the pseudospectral radius of the central cloud at the gap $\rho_{\varepsilon}$ of Fig.~\ref{structured}(c) for different values of $\varepsilon \in [10^{-6}, \;\; 10^{-4}]$.

\begin{figure}
    \centering
    \includegraphics[width=0.45\textwidth]{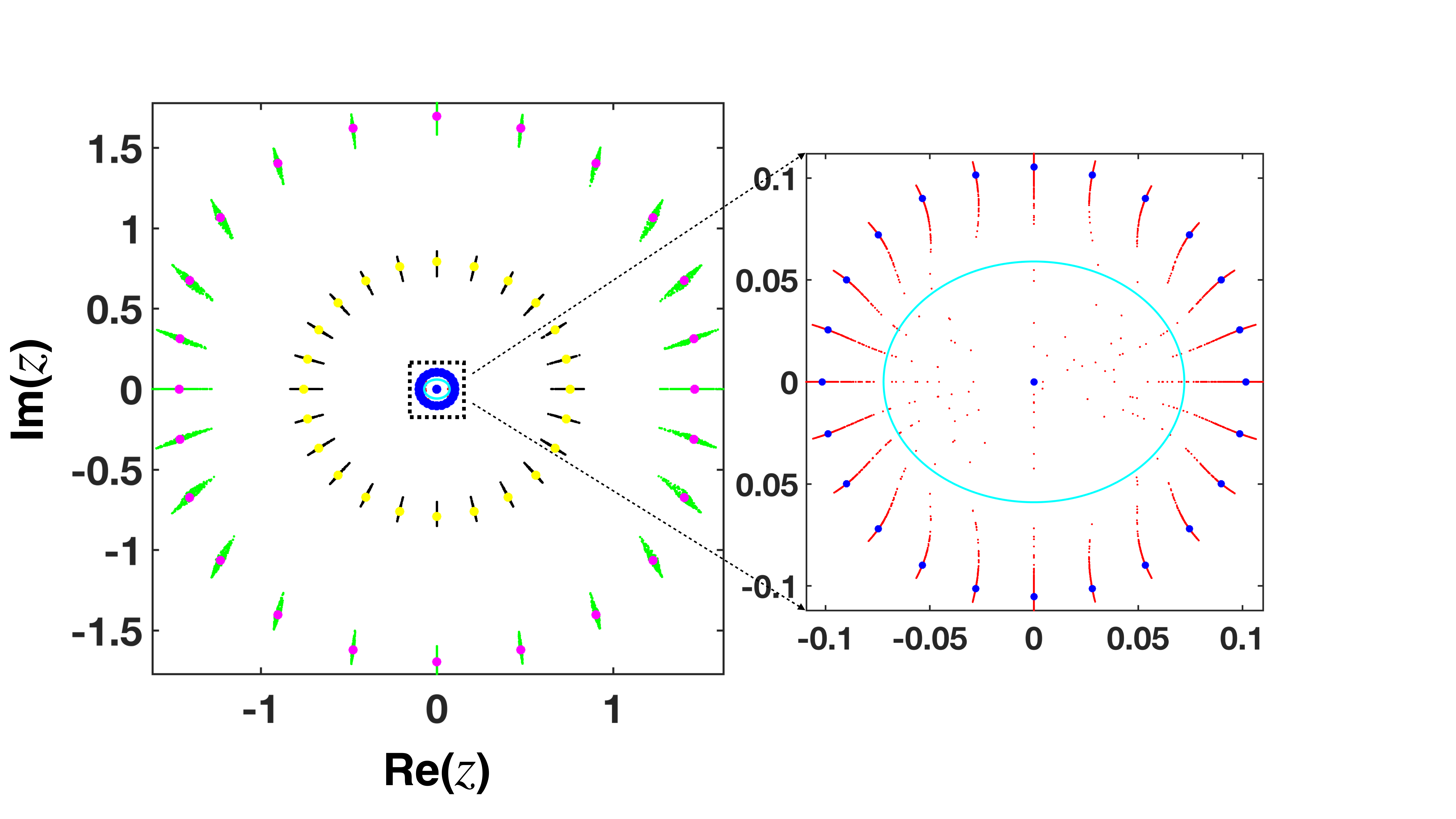}
    \caption{Topological robustness and non-Hermitian sensitivity for chiral perturbations, by locally varying the interface channel's potential strength by nonlinearity. Magenta dots correspond to a lattice with $\gamma = 0.001$, gap size $\approx 4$ and $|\gamma_0| = 1$. Yellow dots are for $\gamma = 1$ with gap size $3.5$ and $|\gamma_0| = 1$. Finally, blue dots are for $\gamma = 2$ with gap size $\approx 0.5$ and $|\gamma_0| = 0.2$. For each of these defect-mode eigenvalues we add perturbations on the coupling coefficients to study their sensitivity. Green, black and red dots denote the corresponding eigenvalue fluctuation for 200 realizations of added perturbation of strength $20\%, 20\%$ and $0.85\%$ respectively. Inset depicts a magnified view for the non-Hermitian case with $\gamma = 2$.}
    \label{robust}
\end{figure}

\textit{Single site local perturbations: robustness vs sensitivity of the zero mode.}--Until now we have examined structured or unstructured perturbations that are globally applied on the whole lattice, where the gain/loss amplitude of the interface channel $\gamma_0$ is equal to zero. It has been recently demonstrated experimentally \cite{Science}, that by using the photorefractive nonlinear effect, we can control not only the real but the imaginary value of the potential strength of the interface channel. Therefore, by tuning of the nonlinearity we can effectively induce two different lattices, a lossy lattice if the interface channel is lossy ($\gamma_0 > 0$), and a gainy lattice when the interface channel is gainy ($\gamma_0 < 0$) (see  SI). Now we are interested to study the structured ``nonlinear" pseudospectra for chiral perturbations on the coupling coefficients. Our results are shown in in Fig.~\ref{robust}, where we apply perturbations to the couplings for various complex values of the interface channel's strength. Thus, the new coupling constants of our system are, $c_1 ^p = c_1 (1 + \varepsilon)$ and $c_2 ^p = c_2 (1 + \varepsilon)$ where, $p$ stands for ``perturbed". We find that the topological robustness of the zero mode is conserved, if and only if, our system is \PT-symmetric with zero gain/loss amplitude at the interface channel, $\gamma_0 = 0$. In all other cases the robustness is gradually lost as we approach the EP, indicting the EP sensitivity. Notice the robustness of the zero mode is also gradually lost as we move away from the value zero. Surprisingly, the change in the energy seems to have a preferred radial direction in the complex plane.

\textit{Discussion and Conclusions.}-- As a general conclusion, non-Hermiticity has an immediate impact on lattice's sensitivity rather than its topological structure. More specifically, we studied the effect of symmetries of the applied perturbations on the overall sensitivity of various lattices. In particular, the complex-unstructured and diagonal-structured ($\mathcal{PT}$ or not) pseudospectra describe the enhanced sensitivity (algebraic root dependence) around the EP's. On the other hand, the topological robustness of the zero mode is revealed only by the structured pseudospectra. In other words, below the HEP's (non-zero gap), the chiral structured perturbations uncover the topological protection of the zero mode. Exactly at the HEP's (zero gap) the situation is more complex. Thus depending on the modal content of the HEP (meaning if the zero mode participates on the EP or not) the topological state is less sensitive on the infinite and finite NHSSH lattices, but it is the most sensitive case of the interface NHSSH. Furthermore, the chiral symmetry of the applied perturbation matrix $E$, leads to reduction of the apparent sensitivity by one order (form EP3 to EP2, in the interface lattice), when the lattice is exactly at the EP3. Finally, we have examined the nonlinearly controlled pseudospectrum of the zero mode, where sensitivity and memory of the topological robustness co-exist. Our results highlight for the first time the fundamental question of the interplay between ultra sensitivity and topological protection in the unique framework of pseudospectra theory and may provide insight for the study of other lattices of non-Hermitian topological physics.\\

This research is supported by the National Key R\&D Program of China under Grant No. 2017YFA0303800, the National Natural Science Foundation (12134006) in China. H.B. acknowledge support in part by the Croatian Science Foundation Grant No. IP-2016-06-5885 SynthMagIA, and the QuantiXLie Center of Excellence, a project co-financed by the Croatian Government and European Union through the European Regional Development Fund - the Competitiveness and Cohesion Operational Programme (Grant KK.01.1.1.01.0004).

\section{References}
\bibliographystyle{longbibliography}

\end{document}